\documentclass[prl,aps,amsfonts,preprint,floats,tightenlines,floatfix]{revtex4}
\usepackage{ifthen}                   
\usepackage{exscale}                  
\usepackage[intlimits]{amsmath}       
\usepackage{amsfonts}
\usepackage{amssymb,amscd}
\usepackage[dvips]{epsfig}
\usepackage{array}
\usepackage{wrapfig}

\newcommand{\be}[1]{\begin{equation} \label{(#1)}}
\newcommand{\ee}{\end{equation}}
\newcommand{\ba}[1]{\begin{eqnarray} \label{(#1)}}
\newcommand{\ea}{\end{eqnarray}}

\begin{document}
\title{{\small \rm {UNITU-THEP-27/2002 \hfill FAU-TP3-02/28}}\\~\\
\large \bf Kugo--Ojima confinement and QCD Green's functions in covariant 
gauges\footnote{Talk given by R.A.\ at the International School on Nuclear 
Physics ``Quarks in Hadrons and Nuclei'' in Erice (Italy), 
September 16 - 24, 2002.}
} 
\author{ R.\ Alkofer, C.\ S.\ Fischer}
\address{Institute for Theoretical Physics, University of T\"ubingen,
          Auf der Morgenstelle 14, D-72076 T\"ubingen}
\author{and L.~von Smekal}
\address{Institute for Theoretical Physics III,
University of Erlangen-N\"urnberg,  Staudtstr.~7, D-91058 Erlangen}

\begin{abstract}

In Landau gauge QCD the Kugo--Ojima confinement criterion and 
its relation to  the infrared behaviour of the gluon and ghost propagators 
are reviewed. It is demonstrated that the realization of this confinement
criterion (which is closely related to the Gribov--Zwanziger horizon condition) 
results from quite general properties of the ghost Dyson--Schwinger equation. 
The numerical solutions for the gluon and ghost propagators obtained from a 
truncated set of Dyson--Schwinger equations provide an explicit example for 
the anticipated infrared behaviour. The results are in good agreement, also 
quantitatively, with corresponding lattice data obtained recently.
The resulting running coupling approaches a fixed point in the infrared, 
$\alpha(0) = 8.915/N_c$. 
Solutions for the coupled system of Dyson--Schwinger equations for the
quark, gluon and ghost propagators are presented. Dynamical generation
of quark masses and thus spontaneous breaking of chiral symmetry takes place.
In the quenched approximation the quark propagator functions agree well with
those of corresponding lattice calculations. For a small number of light 
flavours the quark, gluon and ghost propagators deviate only slightly from
the ones in quenched approximation.
While the positivity violation of the gluon
spectral function is manifest in the gluon propagator,
there are no clear indications of analogous 
positivity violations for quarks so far.

\end{abstract}

\maketitle

\vskip 5mm

\section{Gluon confinement in Landau gauge and\\ the infrared behaviour of the 
ghost propagator}
\label{sec:1}

``Quarks in Hadrons and Nuclei'' is the title of this school. Due to the
overwhelming success of perturbative QCD for hadronic reactions at high
energies we are all convinced that QCD is the correct theory of strong
interactions and that all hadrons are made of quarks and the particles gluing
them together, the gluons. As, however, no quarks and gluons have been seen in
detectors we need the hypothesis of confinement in order to rescue the success
of QCD. Over the last decades there have been many attempts to prove
confinement from QCD. Despite these efforts it is fair to say that the
phenomenon of confinement is still little understood: a clear and undisputable
mechanism responsible for this effect has not been found yet. Furthermore, we
may even face the challenge that nowadays' formulation of quantum field theory
is not sufficient to tackle this problem successfully: it seems not even clear,
at present, whether the phenomenon of confinement is at all compatible with a
description of quark and gluon correlations in terms of local fields.

On the other hand, there is a number of criteria which signal unambigously the
occurrence of confinement. One line of research starts from the expectation
that the two-point correlation functions of QCD, the quark, gluon and ghost
propagators, are likely to provide some clues to the underlying structures of the
theory which are responsible for confinement. And indeed, it has been argued
\cite{Kugo:1995km} that in Faddeev--Popov quantized Landau gauge QCD the
infrared behaviour of the ghost propagator is related to both, the Kugo--Ojima
confinement criterium \cite{Kugo:1979gm} and the Gribov--Zwanziger horizon
condition \cite{Gribov:1978wm,Zwanziger:1993qr}. 

According to Kugo and Ojima \cite{Kugo:1979gm} a physical state space that
only contains colourless states is generated, if two conditions are fulfilled:
First, one should not have massless particle poles in transverse gluon
correlations and, second, one needs well-defined, i.e.\ unbroken, global
colour charges. The second condition can be related to the behaviour of the
ghost propagator in Landau gauge.  For it to be satisfied, the propagator must
be more singular than a massless  particle  pole in the infrared
\cite{Kugo:1995km}.

Gribov's horizon condition is connected to the gauge fixing ambiguities in the
linear covariant gauge \cite{Gribov:1978wm}. Ideally one would eliminate Gribov
copies along gauge orbits by a restriction of the functional integral of the
QCD partition function to the so-called fundamental modular region. This part
of configuration space lies inside the first Gribov region, a convex region in
gauge field space which contains the trivial configuration $A\equiv 0$. At the
boundary of the first Gribov region, the lowest eigenvalue of the 
Faddeev--Popov operator approaches zero. Entropy arguments have been employed
to reason that the infrared modes of the gauge field are  close to this Gribov
horizon \cite{Zwanziger:1993qr}. As the ghost propagator  is the inverse of
the  Fadeev--Popov operator we therefore encounter the presence of the Gribov
horizon in the infrared behaviour of the ghost: The ghost propagator is
required  to be more singular than a simple pole if the restriction to the
Gribov  region is correctly implemented. Furthermore, by the same entropy
arguments, the gluon propagator has to vanish in the infrared
\cite{Zwanziger:1993qr}.

There is an interesting point to note in this context: Employing {\it
Stochastic Quantiziation} instead of the Faddeev--Popov formalism avoids  the
Faddeev--Popov determinant and thus the  Gribov problem completely.  The
Faddeev--Popov ghosts being absent the above picture seems to be impossible to
be realized. Nevertheless one finds essentially the same infrared behaviour for the
propagator of the transverse gluons whereas the longitudinal gluons (which are
absent in Faddeev--Popov--Landau gauge) take over the role of the ghosts
\cite{Zwanziger:2002ia}. Therefore the following generic picture in covariant 
gauges seems likely: Negative metric states like ghosts and/or longitudinal
gluons are long-ranged whereas the propagator for transverse gluons vanishes for
long distances. 

Thus there is compelling evidence that an infrared enhanced propagator for
ghosts (or longitudinal gluons) leads to an infrared vanishing  (or, at least,
infrared finite) tranverse gluon propagator. Such a gluon propagator leads to
violation of positivity in the spectral function for transverse gluons (see
e.g.\ Chapter 5 of ref.\ \cite{Alkofer:2000wg} for a review) and thus describes
confined transverse gluons. Speaking somewhat sloppily: In Landau gauge QCD the
gluons are confined by the Faddeev--Popov ghosts which are the long-range 
correlations of the theory.  

\section{Verifying the Kugo-Ojima Confinement Criterion}
\label{sec:2}

\begin{figure}
\begin{center}
\epsfig{file=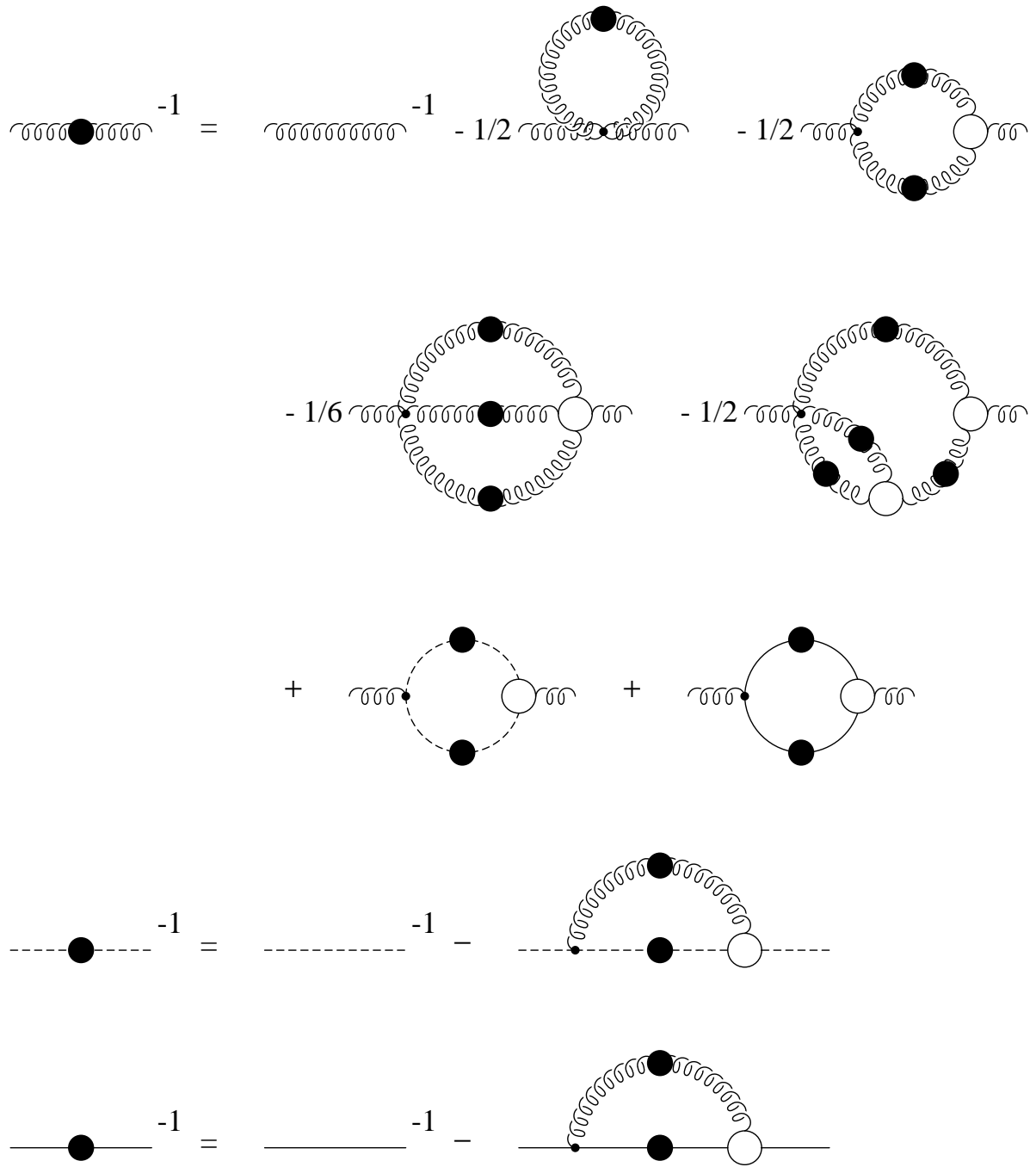,width=160mm}
\end{center}
\caption{Diagrammatic representation of the quark, gluon and ghost 
Dyson--Schwinger equations. 
}
\label{fig:1}       
\end{figure}

As we have argued that confinement in covariant gauges is correlated with
infrared singularities we have the need for a continuum-based non-perturbative
method. The framework we have chosen to investigate the behaviour of the
propagators of QCD are the Dyson--Schwinger equations (DSEs) for the QCD
propagators, see Fig.~\ref{fig:1} (for recent reviews see e.g.\   refs.\
\cite{Alkofer:2000wg,Roberts:2000aa}).  Being complementary to lattice Monte
Carlo simulations which have to deal with finite-volume effects, DSEs  allow
for analytical investigations of the infrared behaviour of correlation
functions. In Landau gauge we have the particularly simple situation that the
ghost-gluon vertex does not suffer from ultraviolet infinities. Based on this
observation one can use the general structure of the ghost DSE, the properties
of multiplicative renormalizability and the assumption that all involved
Green's functions can be expanded in a power series to show that  the
Kugo--Ojima criterion as well as the Gribov--Zwanziger horizon condition are
satisfied \cite{Watson:2001yv,Lerche:2002ep}. Furthermore, it has been shown
that the infrared behaviour of the ghost and the gluon propagators are uniquely
related: Defining ghost and gluon renormalization functions, $Z(k^2)$ and
$G(k^2)$, respectively, from the propagators
$D_{\mu \nu}^{\mbox{\tiny Gluon}}(k^2)=
\left(\delta_{\mu \nu} - \frac{k_\mu k_\nu}{k^2}\right)
{Z(k^2)}/{k^2}$ and $D^{\mbox{\tiny Ghost}}(k^2)= - {G(k^2)}/{k^2}$,
one obtains:
\begin{equation}
Z(k^2) \sim (k^2)^{2\kappa} \qquad {\rm and} \qquad G(k^2) \sim (k^2)^{-\kappa}
. \label{EqExp}
\end{equation}
In addition, the inequality $1/2 \le \kappa < 1$ has been proven
\cite{Lerche:2002ep}. The corresponding gluon propagator is thus infrared
vanishing or, at least, infrared finite.

A further interesting consequence is the fact that  the corresponding powers in
the running coupling (as extracted from the  ghost-gluon vertex) exactly cancel
and one obtains an infrared fixed point for the coupling, see the next section.

\eject ~ \newpage

\section{Propagators of Yang--Mills theory}
\label{sec:3}

As we have seen we can deduce the qualitative infrared behaviour of QCD
propagators applying only general principles. To obtain detailed information on
the propagators of Landau gauge QCD from the DSEs they have to be truncated,
and, even more severe, {\it ans\"atze} for the vertices have to be made.  The
resulting closed system of equations can be solved both, analytically in the 
infrared and numerically for non-vanishing momenta. The considerations
presented in the previous section suggest that for small momenta the ghost loop
dominates in the gluon DSE. Assuming this dominance, effects from a wide class
of possible dressings for the  ghost-gluon vertex have been investigated in
ref.\ \cite{Lerche:2002ep}  and found  to be of negligible influence to the
qualitative findings. 
Thus, for the purpose of this talk we concentrate on the simplest of these
truncation schemes which has been  developed in detail in 
refs.~\cite{Fischer:2002eq,Fischer:2002hn}.  
This scheme employs bare three-point functions and neglects four-gluon vertices. 
In addition, as confinement is expected to be present in the pure Yang--Mills 
sector of QCD we will couple in the quarks at a later stage. 

A coupled system of gluon and ghost DSEs has been studied for the first time in
ref.\ \cite{vonSmekal:1997is}. In this investigation the three-point functions 
have been modeled such that the Slavnov--Taylor identities have been fulfilled
to high degree of accuracy. On the other hand, technical simplifications like
approximating the angular integrals in the DSEs had to be employed. A study
beyond Landau gauge is given in ref.\ \cite{Alkofer:2002}. There it is shown 
that the diagrams involving four-gluon vertices cannot be neglected in the
analytical extraction of the infrared behaviour of the gluon and ghost
propagators in the so-called Curci--Ferrari gauges if bare vertex functions are
used. As a side result it has been shown that in linear covariant gauges the 
assumption of infrared dominance of the ghost loop is, at least,
self-consistent.   

The truncation scheme of refs.\ \cite{Fischer:2002eq,Fischer:2002hn} provides 
the correct one-loop anomalous dimensions of the ghost and  gluon dressing 
functions,
$G(k^2)$ and $Z(k^2)$, respectively, and thus correctly describes the leading
logarithmic behaviour of the  propagators in the ultraviolet. Furthermore, this
scheme reproduces the infrared exponents found in refs.\
\cite{Lerche:2002ep,Zwanziger:2001kw}: 
$$Z(k^2) \sim (k^2)^{2\kappa} \qquad {\rm and} \qquad 
G(k^2) \sim (k^2)^{-\kappa} \qquad {\rm  with} \qquad
\kappa =(93-\sqrt{1201})/98 \approx 0.595.$$ 
These exponents are close to the ones extracted from lattice 
calculations~\cite{Bonnet:2000kw,Bonnet:2001uh,Langfeld:2001cz}. 
Interestingly enough they
are also close to the ones obtained in a comparable truncation scheme in
stochastically quantized Landau gauge Yang--Mills theory for the transverse and
the longitudinal gluons \cite{Zwanziger:2002ia}. 

\begin{figure*}
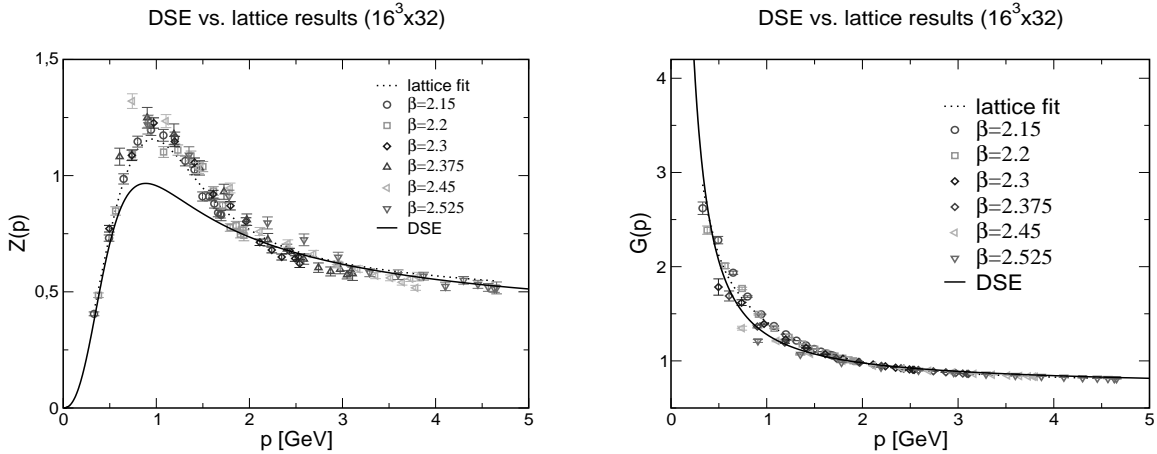

\begin{center}
\epsfig{file=lattice_gluecont.eps,width=7cm,height=6cm}
\hspace{1.0cm}
\epsfig{file=lattice_ghostcont.eps,width=7cm,height=6cm}
\caption{\label{fig:2}Solutions of the Dyson--Schwinger equations 
(labeled DSE) compared to recent lattice results for $N_c=2$
\protect\cite{Langfeld:2001cz}.} 
\end{center}      
\end{figure*}

In Fig.~\ref{fig:2} the numerical solutions for the gluon and ghost dressing
functions for the colour group SU(2) are compared to those obtained from recent
lattice calculations  \cite{Langfeld:2001cz}. Differences mainly occur for the
gluon dressing function in the region around its maximum, i.e.\ somewhat below
one GeV. These can be attributed to the omission of the two-loop diagrams in
the DSE truncation. Given the limitations of both methods the qualitative and
partly even quantitative  agreement is remarkable. 

\newpage

\section{Running coupling}
\label{sec:4}

\begin{figure}
\vspace{0.5cm}
\centerline{
\epsfig{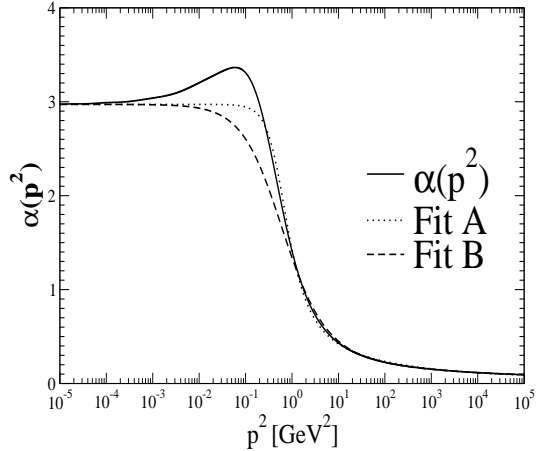}}
\caption{The strong running coupling from the DSEs and the fits 
A and B, c.f.\ eqs.\ (\protect{\ref{fitA},\ref{fitB}}).}
\label{fig:3}       
\end{figure}

A possible non-perturbative definition of the running coupling 
can be given as follows \cite{vonSmekal:1997is,Alkofer:2002ne}:
\begin{equation}
\alpha(k^2) =  \alpha(\mu^2) Z(k^2;\mu^2)G^2(k^2;\mu^2) 
\end{equation}
where the dependence of the propagator functions on the renormalisation point 
have been made explicit.  An important point to notice in the results described
above is the unique relation between the gluon and ghost infrared behaviour.
As explained the structure of the ghost DSE and the non-renormalization of the
ghost-gluon vertex require that in Landau gauge the product $Z(k^2)G^2(k^2)$
goes to a constant in the infrared. The DSE result for the running coupling 
can be seen in Fig.~\ref{fig:3}. 

The analytically obtained value for the fixed  point of the running coupling 
in the infrared is 
$$
\alpha(0)=\frac{4 \pi}{6N_c}
\frac{\Gamma(3-2\kappa)\Gamma(3+\kappa)\Gamma(1+\kappa)}{\Gamma^2(2-\kappa)
\Gamma(2\kappa)} \approx 2.972$$ 
for the gauge
group SU(3) in this truncation scheme. Corrections from possible dressings for
the ghost-gluon vertex have been found to be such that $2.5 < \alpha(0) \le
2.97$ \cite{Lerche:2002ep}. The maximum at non-vanishing momenta
seen in our result for the running coupling results in a multi-valued
beta-function. On the other hand, it appears in a region where the above
comparison to lattice data suggests that our results are least reliable.
(The physical scale has been fixed by requiring the
experimental value $\alpha(M_Z^2=(91.2 \mbox{GeV})^2) = 0.118$.)
We therefore summarize our result for the running coupling
in the monotonic fit functions \cite{Alkofer:2002ne}  
\begin{eqnarray}
\mbox{Fit A:} \quad
\alpha(x) &=& \frac{\alpha(0)}{\ln(e+a_1 (x/\Lambda^2)^{a_2}+
b_1(x/\Lambda^2)^{b_2})}
\label{fitA}\\
\mbox{Fit B:} \quad
\alpha(x) &=& \frac{1}{a+(x/\Lambda^2)^b} 
\Biggl( a \: \alpha(0) + 
\left(\frac{1}{\ln(x/\Lambda^2)}
- \frac{1}{x/\Lambda^2 -1}\right)(x/\Lambda^2)^b\Biggr) 
\label{fitB}
\end{eqnarray}
The value $\alpha(0)=2.972=8.915/N_c$ is known from 
the infrared analysis. In both fits the ultraviolet behaviour 
of the solution fixes the scale,  $\Lambda=0.714 \mbox{GeV}$. 
Note that we have employed
a MOM scheme, and thus $\Lambda$ has to be interpreted as
$\Lambda_{MOM}^{N_f=0}$, i.e.\ this scale has the expected magnitude.
Fit A employs the four additional parameters: 
$a_1=1.106$, $a_2=2.324$,
$b_1=0.004$, $b_2=3.169$.
Fit B (which provides a better description in the ultraviolet at the expense
of some deviations at smaller momenta) has only two free parameters:
$a=1.020$, $b=1.052$. 

\section{Propagators of QCD: Ghost, Glue and Quark}
\label{sec:5}

In the quark DSE as well as in the quark loop of the gluon DSE the  quark-gluon
vertex enters, see Fig.~\ref{fig:1}. Very recently lattice results for the
quark-gluon vertex became available \cite{Skullerud:2002ge}. However, at
present the error bars of such simulations are too large to use the lattice
results as guideline in the construction of reliable {\it ans\"atze} for the
quark-gluon vertex. 
Meanwhile, we proceed with assuming that 
the quark-gluon vertex factorizes as follows \cite{Fischer:2002},
\begin{equation}
\Gamma_\nu(q,k) = V_\nu^{abel}(p,q,k) \, W^{\neg abel}(p,q,k),
\label{vertex-ansatz}
\end{equation}
with $p$ and $q$ denoting the quark momenta and $k$ the gluon momentum. The
non-Abelian factor $W^{\neg abel}$ multiplies an Abelian part  $V_\nu^{abel}$,
which carries the tensor structure of the vertex. For the latter we choose 
a construction \cite{Curtis:1990zs} used widely in QED, see e.g.\ ref.\ 
\cite{Pennington:1998cj}.

As can be infered from the Slavnov--Taylor identity for the quark-gluon vertex
$W^{\neg abel}(p,q,k)$ has to contain factors of the ghost renormalization
function $G(k^2)$. Due to the infrared singularity of the latter the effective
low-energy quark-quark interaction is infrared enhanced as compared to the
interaction generated by the exchange of an infrared suppressed gluon.
Therefore the  effective kernel of the quark DSE contains an (integrable)
infrared singularity, for details see ref.\ \cite{Fischer:2002}. Further
constraints imposed on  $W^{\neg abel}(p,q,k)$ are such that (i) the running
coupling as well as the quark mass function are, as required from general
principles, independent of the renormalization point and (ii) the one-loop
anomalous dimensions of all propagators are reproduced. 

\begin{figure}
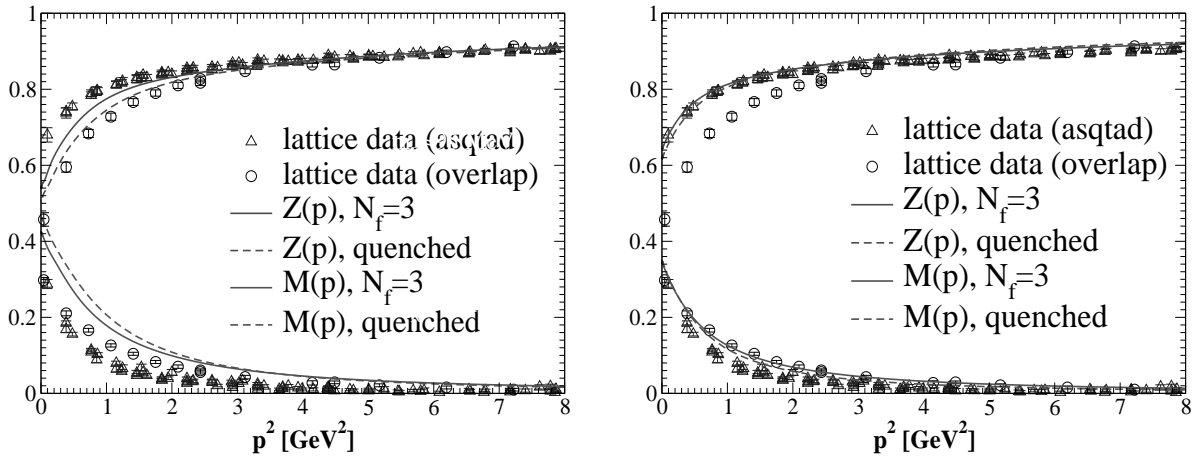

\vspace{0.5cm}
\centerline{
\epsfig{file=quark.d01.eps,width=7.5cm,height=6cm}
\hspace{0.5cm}
\epsfig{file=quark.eps,width=7.5cm,height=6cm}}
\caption{The quark propagator functions in quenched approximation 
as well as for three massless flavours using two different ans{\"atze} for 
the quark-gluon vertex
compared to the lattice data \protect\cite{Bonnet:2002ih}.}
\label{fig:4}       
\end{figure}

In Fig.~\ref{fig:4} we compare our results for the  quark propagator
$\displaystyle S=  \frac{Z(p^2)}{ip\hspace{-.5em}/\hspace{.15em}+M(p^2)}$ in
quenched approximation as well as for three massless flavours with lattice data
\cite{Bonnet:2002ih}. As one sees the DSE results nicely agree with the one 
from the lattice. Furthermore, for the considered number of flavours the
quenched approximation works well. (Also the gluon and the ghost functions
remain almost unchanged \cite{Fischer:2002}.)

\begin{figure}
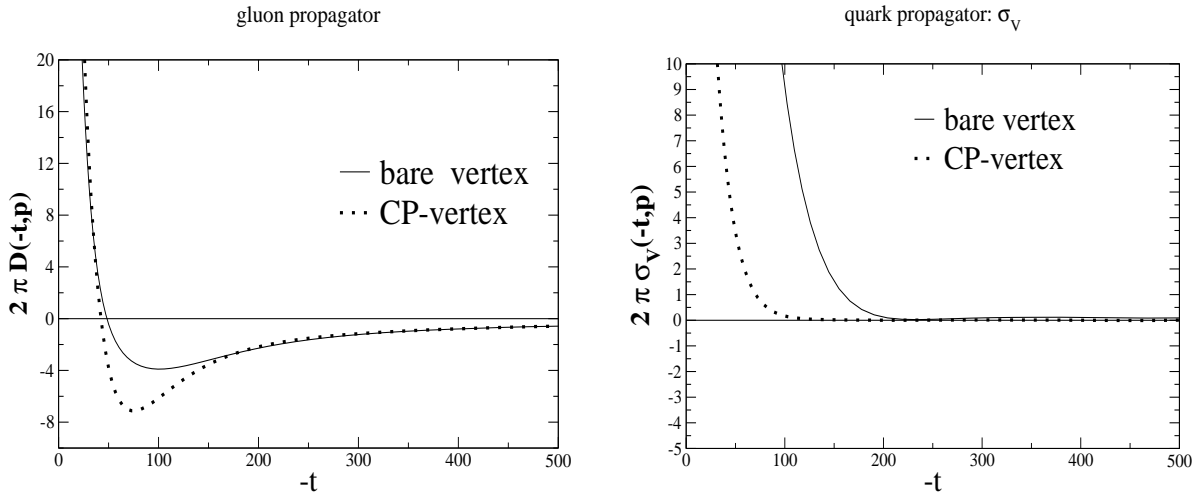

\vspace{0.5cm}
\centerline{
\epsfig{file=pos.glue.eps,width=7.5cm,height=6.5cm}
\hspace{0.5cm}
\epsfig{file=pos.quarkA.eps,width=7.5cm,height=6.5cm}
}
\caption{\label{fig:5} Here we display the one-dimensional Fourier transforms 
of the gluon propagator, $D(-t,\vec{p}^2)$, and the vector part of the 
quark propagator,
$\sigma_V(-t,\vec{p}^2)$.
We observe violation of reflection positivity for the gluon propagator but 
not for the quark propagator.}
\end{figure}

In Fig.~\ref{fig:5} we display the result of a possible test on positivity
violations in the gluon and quark propagators for several choices of the
quark-gluon vertex. Speaking somewhat sloppily, obtaining negative values for
the one-dimensional Fourier transforms of propagators provide a sufficient
condition for positivity violation (and thus confinement). Whereas previous
findings for the gluon propagator are confirmed herewith also beyond quenched
approximation we have not been able to demonstrate positivity violation for
the quark propagator. 

\section{Summary and concluding remarks}
\label{sec:6}

In order to verify the Kugo--Ojima confinement criterion we have studied the 
gluon, ghost and quark Dyson--Schwinger equations of Landau gauge QCD employing
analytical as well numerical techniques. The resulting infrared behaviour of
gluon and ghost propagators, namely a  highly infrared singular ghost and an
infrared suppressed gluon propagator, is related to the Gribov--Zwanziger
horizon condition. The solution for these propagators has then been used to
calculate a non-perturbative running coupling for all spacelike momentum
scales. 
 
The obtained solutions for the quark propagator exhibit dynamical symmetry
breaking. Hereby only carefully constructed vertex {\it ans\"atze} have been
able to generate masses in the typical phenomenological range of $300-400$ MeV.
The agreement with lattice data in quenched approximation confirms the quality
of our truncation and in turn it shows that chiral extrapolation on the lattice
works well. In the unquenched case including the quark-loop in the gluon
equation with $N_f=3$ light quarks we obtain only small corrections compared to
the quenched calculations. For a larger number of light flavours ($N_f>6$)
we have indications that the coupled system is changed qualitatively, and that 
the Kugo--Ojima confinement criterion ceases to be valid.

We looked for positivity violations in the gluon and quark propagators. We
confirmed previous findings that the gluon propagator shows violation of
reflection positivity. We could not find similar violations for the quark
propagator. With the results obtained so far we cannot exclude positivity
violation for the quarks, however, one might take our result as a further
indication that (in Landau gauge) the confinement mechanism for quarks differs
qualitatively from the one for transverse gluons.

Finally we want to remark that studies of the Dyson--Schwinger equations at
non-vanishing temperature (for first results see ref.\ \cite{Maas:2002if})
and density are under way. Main goals are hereby to clarify the relation 
between deconfinement and chiral restauration and to possibly find the
deconfinement criterion related to the Kugo--Ojima criterion.

\section*{Acknowledgements}  
R.A.~thanks  A.\ F\"a\ss ler and 
T.\ Gutsche for organizing  this highly interesting school.\\
We are grateful to K.~Langfeld, J.~M.~Pawlowski, H.~Reinhardt, C.~D.~Roberts,
S.~M.~Schmidt, D.~Shirkov, P.~Tandy, 
P.~Watson and D.~Zwanziger for helpful discussions. 

This work has been supported by the DFG under contract Al 279/3-4 and by the
European graduate school T\"ubingen--Basel (DFG contract GRK 683).


\newpage

\begin{thebibliography}{99}

\bibitem{Kugo:1995km}
T.~Kugo,
Int.\ Symp.\ on BRS symmetry, Kyoto, Sep.~18-22, 1995,
arXiv:hep-th/9511033; \newline
L.~von Smekal and R.~Alkofer,
Proceedings of the 4th International Conference on Quark Confinement
and the Hadron
Spectrum, Vienna, Austria, 3-8 Jul 2000;
arXiv:hep-ph/0009219.


\bibitem{Kugo:1979gm}
T.~Kugo and I.~Ojima,
Prog.\ Theor.\ Phys.\ Suppl.\  {\bf 66} (1979) 1.

\bibitem{Gribov:1978wm}
V.~N.~Gribov,
Nucl.\ Phys.\ B {\bf 139} (1978) 1.

\bibitem{Zwanziger:1993qr}
D.~Zwanziger,
Nucl.\ Phys.\ B {\bf 364} (1991) 127; Nucl.\ Phys.\ B {\bf 399} (1993) 477;
Nucl.\ Phys.\ B {\bf 412} (1994) 657. 

\bibitem{Zwanziger:2002ia}
D.~Zwanziger,
arXiv:hep-th/0206053.

\bibitem{Alkofer:2000wg}
R.~Alkofer and L.~von Smekal,
Phys.\ Rept.\  {\bf 353} (2001) 281
[arXiv:hep-ph/0007355].

\bibitem{Roberts:2000aa}
C.~D.~Roberts and S.~M.~Schmidt,
Prog.\ Part.\ Nucl.\ Phys.\ {\bf 45} (2000) S1 
[arXiv:nucl-th/0005064].

\bibitem{Watson:2001yv}
P.~Watson and R.~Alkofer,
Phys.\ Rev.\ Lett.\  {\bf 86} (2001) 5239
[arXiv:hep-ph/0102332]; \newline
R.~Alkofer, L.~von Smekal and P.~Watson,
Proceedings of the ECT* 
Meeting on {Dynamical Aspects of the QCD
Phase Transition}, Trento, Italy, March 12-15, 2001,
arXiv:hep-ph/0105142.

\bibitem{Lerche:2002ep}
C.~Lerche and L.~von Smekal,
Phys.\ Rev.\ D {\bf 65} (2002) 125006
[arXiv:hep-ph/0202194].

\bibitem{Fischer:2002eq}
C.~S.~Fischer, R.~Alkofer and H.~Reinhardt,
Phys.\ Rev.\ D {\bf 65} (2002) 094008
[arXiv:hep-ph/0202195].

\bibitem{Fischer:2002hn}
C.~S.~Fischer and R.~Alkofer,
Phys.\ Lett.\ B {\bf 536} (2002) 177
[arXiv:hep-ph/0202202].

\bibitem{vonSmekal:1997is}
L.~von Smekal, R.~Alkofer and A.~Hauck,
Phys.\ Rev.\ Lett.\  {\bf 79} (1997) 3591 
[arXiv:hep-ph/9705242];
L.~von Smekal, A.~Hauck and R.~Alkofer,
Annals Phys.\  {\bf 267} (1998) 1
[arXiv:hep-ph/9707327].

\bibitem{Alkofer:2002}
R.~Alkofer, C.~S.~Fischer, H.~Reinhardt and L.~von Smekal,
in preparation.

\bibitem{Zwanziger:2001kw}
D.~Zwanziger,
Phys.\ Rev.\ D {\bf 65} (2002) 094039
[arXiv:hep-th/0109224].

\bibitem{Bonnet:2000kw}
F.~D.~Bonnet, P.~O.~Bowman, D.~B.~Leinweber and A.~G.~Williams,
Phys.\ Rev.\ D {\bf 62} (2000) 051501
[arXiv:hep-lat/0002020].

\bibitem{Bonnet:2001uh}
F.~D.~Bonnet, P.~O.~Bowman, D.~B.~Leinweber, A.~G.~Williams and J.~M.~Zanotti,
Phys.\ Rev.\ D {\bf 64} (2001) 034501
[arXiv:hep-lat/0101013].

\bibitem{Langfeld:2001cz}
K.~Langfeld, H.~Reinhardt and J.~Gattnar,
Nucl.\ Phys.\ B {\bf 621} (2002) 131
[arXiv:hep-ph/0107141];
\newline see also: 
K.~Langfeld {\it et al.},
arXiv:hep-th/0209173.

\bibitem{Alkofer:2002ne}
R.~Alkofer, C.~S.~Fischer and L.~von Smekal,
Acta Phys.\ Slov.\  {\bf 52} (2002) 191
[arXiv:hep-ph/0205125]; 
arXiv:hep-ph/0209366.

\bibitem{Skullerud:2002ge}
J.~Skullerud and A.~Kizilersu,
JHEP {\bf 0209}, 013 (2002)
[arXiv:hep-ph/0205318].


\bibitem{Fischer:2002}
C.~S.~Fischer and R.~Alkofer,
arXiv:hep-ph/0301094.

\bibitem{Curtis:1990zs}
D.~C.~Curtis and M.~R.~Pennington,
Phys.\ Rev.\ D {\bf 42}, 4165 (1990).

\bibitem{Pennington:1998cj}
M.~R.~Pennington,
in {\em Proceedings of the Workshop on Nonperturbative Methods in
Quantum Field Theory}, edited by A.~W. Schreiber, A.~G. Williams, and A.~W.
Thomas, p.~49, Adelaide, 1998, World Scientific;
arXiv:hep-th/9806200.

\bibitem{Bonnet:2002ih}
F.~D.~Bonnet {\it et al.}, 
                  [CSSM Lattice collaboration],
Phys.\ Rev.\ D {\bf 65}, 114503 (2002)
[arXiv:hep-lat/0202003].
J.~B.~Zhang {\it et al.}, 
arXiv:hep-lat/0208037.
P.~O.~Bowman, U.~M.~Heller, D.~B.~Leinweber and A.~G.~Williams,
arXiv:hep-lat/0209129.

\bibitem{Maas:2002if}
A.~Maas, B.~Gruter, R.~Alkofer and J.~Wambach,
arXiv:hep-ph/0210178.

\end{thebibliography}
\end{document}